\documentclass[12pt]{article}
\usepackage{amsmath} 
\input{epsf} 
\def\slash#1{\ooalign{$\hfil/\hfil$\crcr$#1$}} 

\def\ltap{\raisebox{-.6ex}{\rlap{$\,\sim\,$}} \raisebox{.4ex}{$\,<\,$}}

\newcommand\as{\alpha_{\mathrm{S}}} 
\newcommand\f[2]{\frac{#1}{#2}} 
\def\beq{\begin{equation}} 
\def\eeq{\end{equation}} 
\def\beeq{\begin{eqnarray}} 
\def\eeeq{\end{eqnarray}} 
 
\def\to{\rightarrow}
\def\ito{\leftarrow} 
\def\nn{\nonumber}

\def\msbar{{\overline {\rm MS}}} 
 
\def\ptmin{p^l_{T{\rm min}}}
\def\ptmax{p^l_{T{\rm max}}}
\def\ptWW{p_{T}^{WW}}

\def\WW{WW}
\def\tL{{\tilde L}}

\begin{document} 

\begin{titlepage}
\renewcommand{\thefootnote}{\fnsymbol{footnote}}
\begin{flushright}
hep-ph/0510337
     \end{flushright}
\par \vspace{10mm}

\begin{center}
{\Large \bf
Soft-gluon effects\\
\vskip .3cm
in $\WW$ production at hadron colliders}
\end{center}
\par \vspace{2mm}
\begin{center}
{\bf Massimiliano Grazzini}\\

\vspace{5mm}

INFN, Sezione di Firenze\\ and Dipartimento di Fisica,
Universit\`a di Firenze,\\ 
I-50019 Sesto Fiorentino, Florence, Italy\\

\vspace{5mm}

\end{center}

\par \vspace{2mm}
\begin{center} {\large \bf Abstract} \end{center}
\begin{quote}
\pretolerance 10000

We consider QCD radiative corrections to $\WW$ pair production
in hadron collisions.
We perform a calculation that consistently combines next-to-leading order predictions
with soft-gluon resummation valid at small transverse momenta $\ptWW$ of the $\WW$ pair.
We present results for the $\ptWW$ distribution at the LHC up to (almost)
next-to-next-to-leading logarithmic accuracy, and study the effect of resummation on the charged-lepton distributions.
Soft-gluon effects are typically mild,
but they can be strongly enhanced when hard cuts are applied.
The relevant distributions are generally well described by the MC@NLO event generator.

\end{quote}

\vspace*{\fill}
\begin{flushleft}
October 2005

\end{flushleft}
\end{titlepage}

\setcounter{footnote}{1}
\renewcommand{\thefootnote}{\fnsymbol{footnote}}

\section{Introduction}
\label{sec:intro}

Vector boson pair production in hadron collisions is important for 
physics within and beyond the Standard Model (SM).
First of all, this process can be used to measure the vector boson trilinear couplings.
Any deviation from the pattern predicted by $SU(2)\otimes U(1)$ gauge invariance would be a signal of new physics.
The Tevatron collaborations are currently measuring $WW$, $WZ$ and $W\gamma$ cross sections at invariant masses larger than those probed at LEP2,
setting limits on the corresponding anomalous couplings \cite{Elvira:2005ju}.

Furthermore, vector boson pairs are an important background for new physics searches.
If a heavy enough Higgs boson
does exist,
it will
decay with large branching ratios in $\WW$ and $ZZ$ pairs,
whereas charged Higgs bosons from non standard Higgs sectors may decay into $WZ$ final states \cite{Gunion:1989we}.
Typical signals of supersymmetry, e.g. three charged leptons plus missing energy,
find an important background in $WZ$ and $W\gamma$ production \cite{Martin:1997ns}. 

In this context, $\WW$ production has received considerable attention.
This process provides an irreducible background to the Higgs boson search
in the $H\to WW\to l\nu l\nu$ channel, which is the most important
when the mass $M_H$ of the Higgs boson is in the range
$155\ltap M_H\ltap 170$ GeV.
Since two neutrinos are present in the final state, no invariant-mass peak
can be reconstructed.
Fortunately, the angular correlations among the charged leptons
suggest a good discrimination of
the signal over the background \cite{Dittmar:1996ss},
provided the background can be reliably extrapolated into the signal region.
It is thus essential to have good control on the SM prediction for
the $\WW$ cross section as well as for the associated distributions.

QCD corrections to $\WW$ production at next-to-leading-order (NLO)
have been computed more than 10 years ago
\cite{Ohnemus:1991kk,Frixione:1993yp} and enhance the cross section by about $40\%$ at LHC energies.
These calculations were done with the traditional method
of evaluating directly the relevant squared amplitudes.
As a consequence, the $W$'s polarization was summed over
and spin correlations were not taken into account. 

More recent NLO calculations exist that, using the one-loop helicity amplitudes computed in
Ref.~\cite{Dixon:1998py}, fully
take into account
spin correlations \cite{Dixon:1999di}.
The general purpose NLO program MCFM \cite{Campbell:1999ah} includes
in addition single-resonant contributions neglected in the calculation of
Ref.~\cite{Dixon:1999di}.
Very recently, the effect of the potentially large
next-to-next-to-leading (NNLO) contribution
from the one-loop $gg\to WW\to l\nu l\nu$
diagrams has also been studied \cite{Binoth:2005ua,Duhrssen:2005bz}.

The fixed-order NLO calculations provide a reliable estimate of $\WW$ cross sections and distributions
as long as the scales involved in the process are all of the same order.
When the transverse momentum of the $\WW$ pair $\ptWW$ is much smaller
than its invariant mass $M_{WW}$ the validity of the fixed-order expansion may
be spoiled since the coefficients of the perturbative expansion
can be enhanced by powers of the large logarithmic terms, $\ln^n M_{WW}/\ptWW$.
This is certainly the case for the $\ptWW$ spectrum, which, when evaluated at fixed order, is even divergent as $\ptWW\to 0$, and thus requires an all-order
resummation of the logarithmically enhanced terms.
Resummation effects, however, can be visible also in other observables, making it important to study them in detail.

The way to perform transverse-momentum resummation is known \cite{Dokshitzer:hw}-\cite{Catani:2000vq}:
to correctly implement momentum conservation in the transverse plane the resummation
has to be performed in impact parameter $b$-space, $b$ being the variable conjugated to $p_T$ through
a Fourier transformation.
In the case of vector boson pair production, transverse-momentum resummation has been applied to $\gamma\gamma$
\cite{Balazs:1997hv} and $ZZ$ \cite{Balazs:1998bm} pair production.
To our knowledge, the case of $\WW$ pair production has not been considered yet.

In the present paper we report on an implementation of $b$-space
resummation to $\WW$ production in hadron collisions.
We use the helicity amplitudes of Ref.~\cite{Dixon:1998py} and work
in the narrow width approximation (i.e. we only consider double-resonant contributions),
but fully include the decay of the $W$'s, keeping track of their polarization in the leptonic decay.
In the large $\ptWW$ region we use LO perturbation theory ($\WW$+1 parton); in the region $\ptWW\ll M_{WW}$
the large logarithmic contributions are resummed to (almost)
next-to-next-to-leading logarithmic
(NNLL) \cite{deFlorian:2000pr,deFlorian:2001zd} accuracy.
Our results have thus
uniform
NLO accuracy but consistently 
include the effect of resummation in the region $\ptWW\ll M_{WW}$.

To perform the resummation
we use the formalism of Refs.~\cite{Bozzi:2003jy,Bozzi:2005wk}.
In this approach, the resummation is achieved at the level of the partonic
cross section and the large logarithmic contributions are exponentiated
in a process-independent manner, being constrained to give vanishing
contribution to the total cross section.

Besides computing transverse momentum spectra of the $\WW$ pair,
we study
the effect of resummation on the leptonic distributions, when different cuts are applied.
We also compare our results with those obtained at NLO and with the ones from
the general purpose event generator MC@NLO \cite{MCatNLO}, which,
in its latest release \cite{Frixione:2005gz}, partially includes the effect of spin correlations
in the $W$'s decay.

The paper is organized as follows. In Sect.~\ref{sec:theory} we discuss
the application of $b$-space resummation to $\WW$ production.
In Sect.~\ref{sec:res} we present our numerical results
and in Sect.~\ref{sec:concl} we draw our conclusions.

Preliminary results of this work for the $\ptWW$ spectrum
were used in the study of Ref.~\cite{Davatz:2004zg}.

\section{Transverse-momentum resummation for $\WW$ pair production}
\label{sec:theory}

In this Section we apply $b$-space resummation to the production of
$\WW$ pairs in hadron collisions.
The resummation formalism we use
is completely general, and, as
discussed in detail in Ref.~\cite{Bozzi:2005wk}, can be applied to a generic process in which a system
of non strongly interacting particles of high mass $M$ is produced in hadronic collisions.
In the case of $\WW$ production, the mass $M$ is the invariant mass of the $\WW$ system.

The resummation is performed at the level of the partonic cross section, which is decomposed as:
\begin{equation}
\label{resplusfin}
\f{d{\hat \sigma}_{WW\,ab}}{dM^2 dp_T^2}=
\f{d{\hat \sigma}_{WW\,ab}^{(\rm res.)}}{dM^2 dp_T^2}
+\f{d{\hat \sigma}_{WW\,ab}^{(\rm fin.)}}{dM^2 dp_T^2}\, .
\end{equation}
The first term on the right hand side,
$d{\hat \sigma}^{({\rm res.})}_{WW\, ab}$,
contains all the logarithmically enhanced contributions at small $p_T$,
and has to be evaluated by resumming them to all orders in $\as$.
The second term, $d{\hat \sigma}^{({\rm fin.})}_{WW\, ab}$,
is free of such contributions, and
can thus be evaluated at fixed order in perturbation theory. 

The resummed component $d{\hat \sigma}^{({\rm res.})}_{WW\, ab}$
can be expressed as
\begin{equation}
\label{resum}
\f{d{\hat \sigma}_{WW \,ab}^{(\rm res.)}}{d M^2 dp_T^2}(p_T,M,{\hat s};
\as(\mu_R^2),\mu_R^2,\mu_F^2) 
=\f{M^2}{\hat s} \;
\int_0^\infty db \; \f{b}{2} \;J_0(b p_T) 
\;{\cal W}_{ab}^{WW}(b,M,{\hat s};\as(\mu_R^2),\mu_R^2,\mu_F^2) \;,
\end{equation}
where $J_0(x)$ is the $0$-order Bessel function, $\mu_R$ ($\mu_F$) is
the renormalization (factorization) scale and ${\hat s}$ is
the partonic centre-of-mass energy.
By taking the $N$-moments of ${\cal W}$ with respect to the variable $z=M^2/{\hat s}$ at fixed $M$
the resummation structure of ${\cal W}_{ab, \,N}^F$ can indeed be organized in exponential form
\footnote{Here, to simplify the notation, flavour indices are understood, or in other words,
we limit ourselves to discussing the flavour non-singlet contribution. A complete discussion
of the exponentiation structure in the general case can be found in Appendix A of Ref.~\cite{Bozzi:2005wk}.}
\begin{align}
\label{wtilde}
{\cal W}_{N}^{WW}(b,M;\as(\mu_R^2),\mu_R^2,\mu_F^2)
&={\cal H}_{N}^{WW}\left(M, 
\as(\mu_R^2);M^2/\mu^2_R,M^2/\mu^2_F,M^2/Q^2
\right) \nonumber \\
&\times \exp\{{\cal G}_{N}(\as(\mu^2_R),L;M^2/\mu^2_R,M^2/Q^2
)\}
\;\;,
\end{align}
were we have defined the logarithmic expansion parameter $L$ as
\begin{equation}
\label{logpar}
L\equiv \ln \f{Q^2 b^2}{b_0^2}
\end{equation}
and the coefficient $b_0=2e^{-\gamma_E}$ ($\gamma_E=0.5772...$ is the Euler number) has a kinematical origin.
The scale $Q$ appearing in Eqs.~(\ref{wtilde},~\ref{logpar}),
named resummation scale in Ref.~\cite{Bozzi:2005wk},
parametrizes the
arbitrariness in the resummation procedure, 
and has to be chosen of the order of the hard scale $M$.
Variations of $Q$ around $M$ can give an idea of the size of yet uncalculated higher-order
logarithmic contributions.
The function ${\cal H}_N^{WW}$ does not depend on the impact parameter $b$ and it includes all the perturbative
terms that behave as constants as $b\to\infty$. It can thus be expanded in powers of $\as=\as(\mu_R^2)$:
\begin{align}
\label{hexpan}
{\cal H}_N^{WW}(M,\as;M^2/\mu^2_R,M^2/\mu^2_F,M^2/Q^2)&=
\sigma_{WW}^{(0)}(\as,M)
\Bigl[ 1+ \f{\as}{\pi} \,{\cal H}_N^{WW \,(1)}(M^2/\mu^2_R,M^2/\mu^2_F,M^2/Q^2) 
\Bigr. \nn \\
&+ \Bigl.
\left(\f{\as}{\pi}\right)^2 
\,{\cal H}_N^{WW \,(2)}(M^2/\mu^2_R,M^2/\mu^2_F,M^2/Q^2)+\dots \Bigr]\,
\end{align}
where $\sigma_{WW}^{(0)}$ is the Born partonic cross section.
The exponent ${\cal G}_N$ includes the complete dependence on $b$ and, in particular, it contains all
the terms that order-by-order in $\as$ are logarithmically divergent as $b\to\infty$.
The logarithmic expansion of ${\cal G}_N$ reads
\begin{align}
\label{exponent}
{\cal G}_{N}(\as(\mu^2_R),L;M^2/\mu^2_R,M^2/Q^2)&=L g^{(1)}(\as L)+g_N^{(2)}(\as L;M^2/\mu_R^2,M^2/Q^2)\nn\\
&+\f{\as}{\pi} g_N^{(3)}(\as L,M^2/\mu_R^2,M^2/Q^2)+\dots
\end{align}
where the term $L\, g^{(1)}$ collects the LL contributions, the function $g_N^{(2)}$ includes
the NLL contributions, $g_N^{(3)}$ controls the NNLL terms and so forth.

In the implementation of Eq.~(\ref{wtilde}) the resummation of the large logarithmic contributions affects
not only the small-$p_T$ region ($p_T\ll M$), but also the region of large $p_T$ ($p_T\sim M$).
This can be easily understood by observing that the logarithmic expansion parameter $L$ is divergent as $b\to 0$.
To reduce the impact of unjustified higher-order contributions in the large-$p_T$ region,
the logarithmic variable $L$ in Eq.~(\ref{logpar}) is replaced as
\begin{equation}
\label{ltilde}
L\to\tL~~~~~~~~~~~~\tL\equiv \ln \left(\f{Q^2 b^2}{b_0^2}+1\right)\, .
\end{equation}
The variables $L$ and $\tL$ are equivalent when $Qb\gg 1$, but they lead to a different behaviour
of the form factor at small values of $b$ (i.e. large values of $p_T$).
When $Qb\ll 1$ in fact, $\tL\to 0$ and $\exp\{{\cal G}_N\}\to 1$.
The replacement in Eq.~(\ref{ltilde}) has thus a twofold consequence: it reduces the impact of resummation at large values of $p_T$, and
it allows us to recover the corresponding fixed-order cross section upon integration over $p_T$.

Another important property of the formalism of Ref.~\cite{Bozzi:2005wk} is that
the process dependence (as well as the factorization scale and scheme dependence) is fully encoded
in the hard function ${\cal H}^{WW}$.
In other words, the functions $g^{(i)}$ are universal: they depend only on the channel
in which the process occurs at Born level, ($q{\bar q}$ annihilation
in the case of $\WW$ production). Their explicit expressions up to $i=3$
are given in Ref.~\cite{Bozzi:2005wk}
in terms of the universal perturbative coefficients
$A_q^{(1)}$, $A_q^{(2)}$, $A_q^{(3)}$,
${\tilde B}_{q,N}^{(1)}$, ${\tilde B}_{q,N}^{(2)}$.
In particular, the LL function $g^{(1)}$ depends on the coefficient $A_q^{(1)}$, the NLL function $g_N^{(2)}$
also depends on $A_q^{(2)}$ and ${\tilde B}_q^{(1)}$ and the NNLL function $g_N^{(3)}$ also depends on $A_q^{(3)}$ and ${\tilde B}_{q,N}^{(2)}$.
All these coefficients are known except $A_q^{(3)}$.
In our quantitative study (see Sect.~\ref{sec:res}) we assume
that the value of $A_q^{(3)}$
is the same as the one \cite{Vogt:2000ci,Moch:2004pa}
that appears in resummed calculations of soft-gluon contributions near
partonic threshold.

We now turn to the hard coefficients ${\cal H}_{q{\bar q}\ito ab,N}^{WW}$.
The first order
contributions ${\cal H}_{q{\bar q}\ito ab,N}^{WW(1)}$ in Eq.~(\ref{hexpan})
are known.
The flavour off-diagonal part is process independent
and in the $\msbar$ scheme it reads:
\begin{equation}
\label{H1off}
{\cal H}_{q{\bar q}\ito gq,N}^{(1)}={\cal H}_{gg\ito qg,N}^{(1)}
=\f{1}{2(N+1)(N+2)} \, + \gamma_{qg,N}^{(1)}\, \ln\f{Q^2}{\mu_F^2}\, ,
\end{equation}
where $\gamma^{(1)}_{ab,N}$ are the LO anomalous dimensions.
The flavour diagonal coefficient
is instead process dependent, and, as shown in Refs.~\cite{deFlorian:2000pr,deFlorian:2001zd}, 
it can be expressed
in terms of the finite part ${\cal A}^{WW}$
\footnote{We adopt here for ${\cal A}^{WW}$ the definition of
Eq.~(38) of Ref.~\cite{deFlorian:2001zd}.}
of the one-loop correction
to the Born subprocess,
computed in Ref.~\cite{Dixon:1998py}.
We have \cite{Bozzi:2005wk}:
\begin{equation}
\label{H1}
{\cal H}^{WW(1)}_{q{\bar q}\ito q{\bar q},N}=
C_F\left(\f{1}{N(N+1)}+\f{\pi^2}{6}\right)+\f{1}{2} {\cal A}^{WW}
-\left(B^{(1)}_q+\f{1}{2} A^{(1)}_q \ln\f{M^2}{Q^2}\right)\ln\f{M^2}{Q^2}
+2\gamma^{(1)}_{qq}\, \ln\f{Q^2}{\mu_F^2}\, .
\end{equation}
The second order coefficients
${\cal H}_{q{\bar q}\ito ab,N}^{WW(2)}$
in Eq.~(\ref{hexpan}) have not yet been computed.

We finally consider the finite component in Eq.~(\ref{resplusfin}).
This contribution has to be evaluated
starting from the usual perturbative truncation of the partonic cross section
and subtracting the expansion of the resummed part at the {\em same} perturbative order.
This procedure allows us to combine the resummed and the finite component of
the partonic cross section to achieve uniform theoretical accuracy over the entire range
of transverse momenta.
Note that, since for $\WW$ production at $\ptWW\neq 0$ only the LO result is known
($WW$+1 parton), we can perform the matching at LO only.

In summary, the inclusion of the functions $g^{(1)}$, $g_N^{(2)}$
and of the coefficient
${\cal H}_N^{WW(1)}$ in the resummed component,
together with the evaluation of the finite component at LO,
allows us to perform the resummation at NLL+LO accuracy.
The inclusion of the function $g_N^{(3)}$ (still performing the matching at LO)
allows us to reach (almost) NNLL+LO accuracy
\footnote{The reader should not be confused by this notation: the NLL+LO and NNLL+LO
results include the complete ${\cal O}(\as)$ real contribution through
the finite component in Eq.~(\ref{resplusfin})
as well as
the full virtual ${\cal O}(\as)$ correction through the
${\cal H}^{WW(1)}_{q{\bar q}\ito q{\bar q},N}$ coefficient
in Eq.~(\ref{H1}). As a consequence, our NLL+LO and NNLL+LO results 
contain the complete NLO correction plus resummation of the 
large logarithmic contributions at $\ptWW\ll M_{WW}$.}.
The reason why we cannot claim full NNLL accuracy is that
the NNLL contribution $\as g_N^{(3)}$ in the exponent in Eq.~(\ref{exponent})
is of the same order of the terms coming from the
combined effect of $\as^2 {\cal H}_N^{WW(2)}$ and
$Lg^{(1)}$, which are not under control.
For this reason, in the following Section we will mainly rely
on our NLL+LO prediction,
regarding the NNLL+LO result as an indication of the size of NNLL effects.

\section{Results}
\label{sec:res}

In this Section we present numerical results
for $\WW$ production in $pp$ collisions at LHC energies.
We compare our resummed perturbative
predictions at NLL+LO and NNLL+LO accuracy
with the NLO ones, obtained with the general purpose program MCFM \cite{Campbell:1999ah},
and with results obtained with the MC@NLO event generator \cite{Frixione:2005gz}.

To compute the $\WW$ cross section we use MRST2002 NLO densities \cite{Martin:2002aw}
and $\as$ evaluated at two-loop order.
As discussed in Sect.~\ref{sec:theory},
our resummed predictions depend on renormalization, factorization and resummation scales.
Unless stated otherwise, the resummation scale is set equal to the
invariant mass $M_{WW}$ of the $\WW$ pair, whereas
renormalization and factorization scales are set to $2M_W$.
The latter choice allows us to exploit our unitarity constraint and to exactly recover
the total NLO cross section when no cuts are applied.
At NLO we consistently use $\mu_F=\mu_R=2M_W$ as default choice,
whereas in MC@NLO $\mu_F$ and $\mu_R$ are set to the default choice, the average transverse mass of the $W$'s.

The predictions of resummation are implemented in a partonic Monte Carlo program which
generate the full 5-body final state ($l\nu l \nu$ + 1 parton).
Nonetheless, since the resummed cross section in Eq.~(\ref{resplusfin}) is inclusive over rapidity,
we are not able to apply the usual rapidity cuts on the leptons\footnote{Note that this is not
a limitation of principle: the resummation formalism can be extended to the double differential
transverse momentum and rapidity distribution.}.
To the purpose of the present work, we do not expect this limitation to be essential.
Note also that, since the resummation formalism we use is valid for the inclusive production of a non-strongly interacting final state
($WW\to l\nu l\nu$ in the present case), we are not allowed to apply cuts on the accompanying jets.

We start by considering the inclusive cross sections.
Our NLL+LO (NNLL+LO) result is 115.6 (115.5) pb,
and agrees with the NLO one (116.0 pb) to better than $1\%$.
The cross section from MC@NLO is instead lower, about 114.7 pb. The above difference is due to the different choice
of the scales, and to the different convention in the choice of the electroweak couplings adopted in MC@NLO.
\begin{figure}[htb]
\begin{center}
\begin{tabular}{cc}
\epsfysize=6truecm
\epsffile{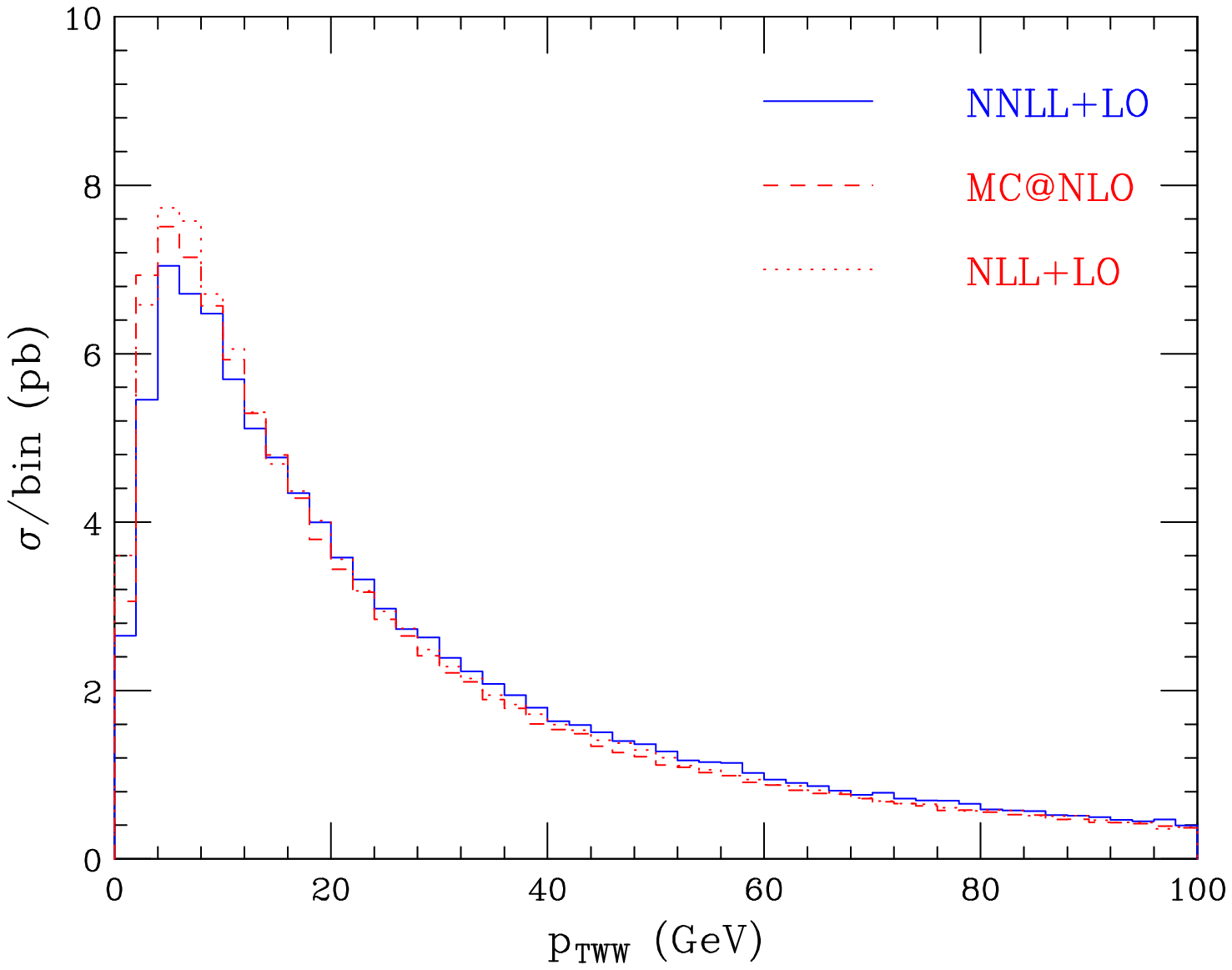} & \epsfysize=6truecm\epsffile{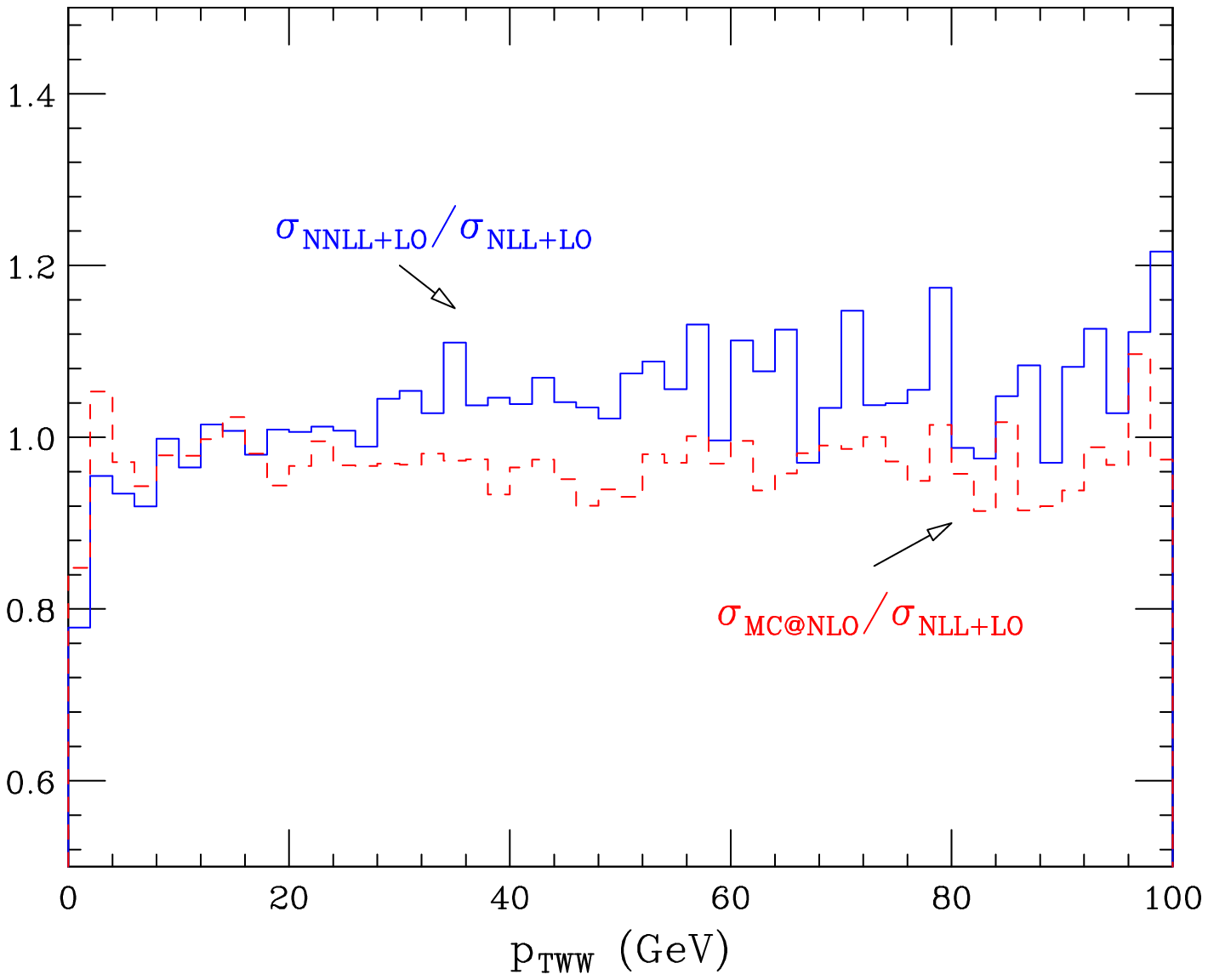}\\
\end{tabular}
\end{center}
\caption{\label{fig:ptww}{\em Left: comparison of the transverse momentum spectra of the $\WW$ pair obtained at NLL+LO, NNLL+LO and with MC@NLO. No cuts are applied. Right: NNLL+LO and MC@NLO results normalized to NLL+LO.}}
\end{figure}

In Fig.~\ref{fig:ptww} (left) we show the corresponding $p_T$ distribution,
computed at NLL+LO (dotted), NNLL+LO (solid) and with MC@NLO (dashed). 
The NLO result, not shown in the plot, diverges to $+\infty$ as $\ptWW\to 0$.
We see that the three histograms are very close to each other and show a peak around $\ptWW\sim 5$ GeV.
The agreement is confirmed in  Fig.~\ref{fig:ptww} (right) where the ratio of the NNLL+LO (solid) and MC@NLO (dashed) results to the NLL+LO are displayed.
We see that, apart from statistical fluctuations, the MC@NLO and NLL+LO agree  within 2-3 $\%$ on average.
On the contrary, the NNLL contribution tends to make the distribution harder and its effect increases at higher $p_T$, always being below $10\%$.

In order to study the perturbative uncertainties affecting our NLL+LO calculation, 
we have varied $\mu_F$ and $\mu_R$ by a factor 2 around the central value.
We find that the effect of scale variations is rather small, of the order of $\pm 1\%$,
and comparable with the estimated accuracy of our numerical code.
Similar results are obtained by varying $\mu_F$ and $\mu_R$ in MC@NLO.
\begin{figure}[htb]
\begin{center}
\begin{tabular}{c}
\epsfysize=8truecm
\epsffile{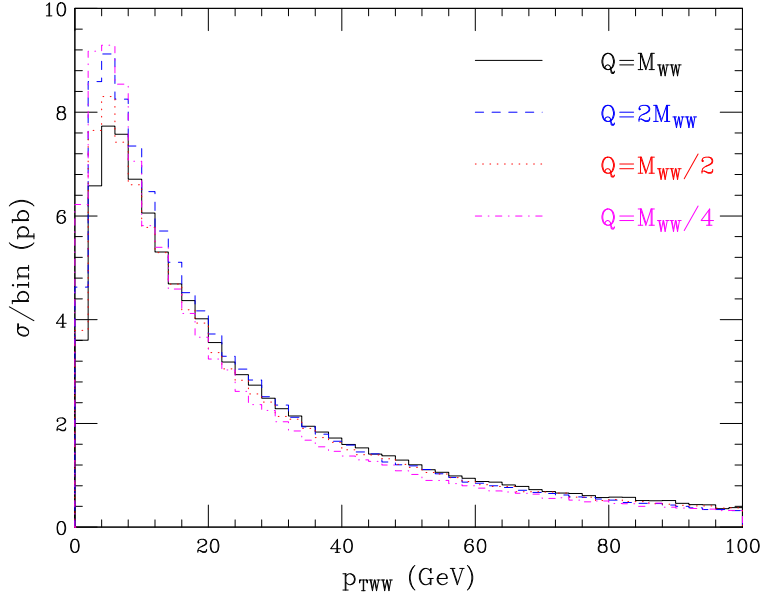}
\end{tabular}
\end{center}
\caption{\label{fig:ptwwadep}{\em NLL+LO spectra for different values of the resummation scale $Q$.}}
\end{figure}

The dependence of our NLL+LO results on the resummation scale is instead stronger.
In Fig.~\ref{fig:ptwwadep} we show the NLL+LO prediction for
different values of the resummation scale $Q$. 
We see that varying the resummation scale the effect on the $\ptWW$ spectrum is well visible
and amounts to about $\pm 10\%$ at the peak.
For lower (higher) values of $Q$ the effect of the resummation is confined to smaller (larger) values of $\ptWW$.
Comparing Figs.~\ref{fig:ptww} and \ref{fig:ptwwadep} we see that the order of magnitude of the
NNLL effect, partially included in
our NNLL+LO prediction, is smaller than the spread in the NLL+LO result from resummation-scale variations.

As in the case of Higgs production \cite{Bozzi:2005wk}, we find that the choice $Q=2M_{WW}$ gives
(slightly) negative cross sections at very large $\ptWW$.
In order to define a range of variation of $Q$,
we would like to avoid values that give a bad behaviour at large $\ptWW$.
For this reason in the following we will consider
resummation-scale variations in the range $M_{WW}/4 \leq Q\leq M_{WW}$.
\begin{figure}[htb]
\begin{center}
\begin{tabular}{c}
\epsfysize=8truecm
\epsffile{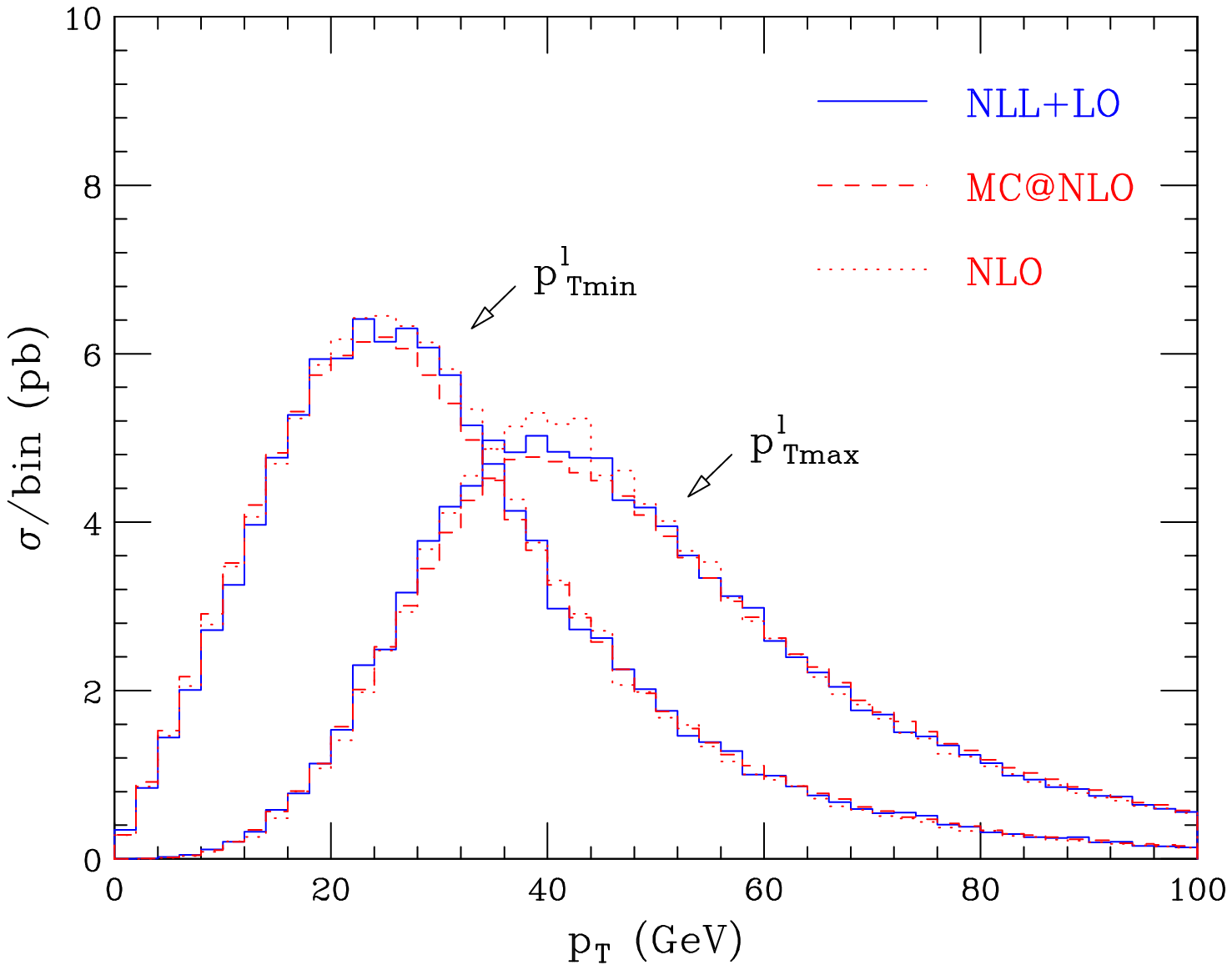}
\end{tabular}
\end{center}
\caption{\label{fig:ptlept}{\em Distributions in $\ptmin$ and $\ptmax$. No cuts are applied.}}
\end{figure}

We now consider the $p_T$ spectra of the leptons.
For each event, we classify the transverse momenta of the two charged leptons
into their minimum and maximum values, $\ptmin$ and $\ptmax$.
In Fig.~\ref{fig:ptlept} we plot the corresponding $p_T$ spectra,
computed at NLL+LO (solid), NLO (dotted) and with MC@NLO (dashes).
We see that all the three predictions are in good agreement.
Small differences are visible in the peak of the $p_T$ distribution of the lepton
with larger $p_T$: the peak predicted by MC@NLO and by the NLL+LO calculation is slightly lower than the one from the NLO calculation.
The renormalization- and factorization-scale
dependence of the results, defined as above,
is still of about $\pm 1\%$.
The scale uncertainty of the NLL+LO result is
dominated by resummation-scale variations
and it is about $\pm 2-3\%$ on average.

We then examine the impact of resummation when cuts on the final state are applied.
We start by defining the following selection criteria, taken from the study of Ref.~\cite{Davatz:2004zg}.

{\bf Cuts A:}
\begin{itemize}
\item The $p_T$ of the charged leptons should be larger than $20$ GeV.
\item The invariant mass $m_{ll}$ of the charged leptons should be smaller than $80$ GeV.
\item The missing $p_T$ of the event should be larger than $20$ GeV. 
\item The azimuthal separation $\Delta\phi$ of the charged leptons in the transverse plane should be smaller than $135^o$.
\end{itemize}
These cuts basically select a pair of $W$'s, suppressing events with lepton pairs originating from the decay of a $Z$.
The corresponding NLL+LO cross section is 21.03 pb, about $5\%$ smaller than the NLO result of 22.10 pb. The MC@NLO cross section is instead 21.16 pb.
The differences between these results are not unexpected: both the NLL+LO and the MC@NLO calculations enforce a unitarity constraint which ensure
the correct NLO normalization is recovered
when {\em total} cross sections are considered.
When, as in this case, cuts are applied, higher order effects are present that make NLL+LO and MC@NLO results generally different from NLO.
\begin{figure}[htb]
\begin{center}
\begin{tabular}{c}
\epsfysize=8truecm
\epsffile{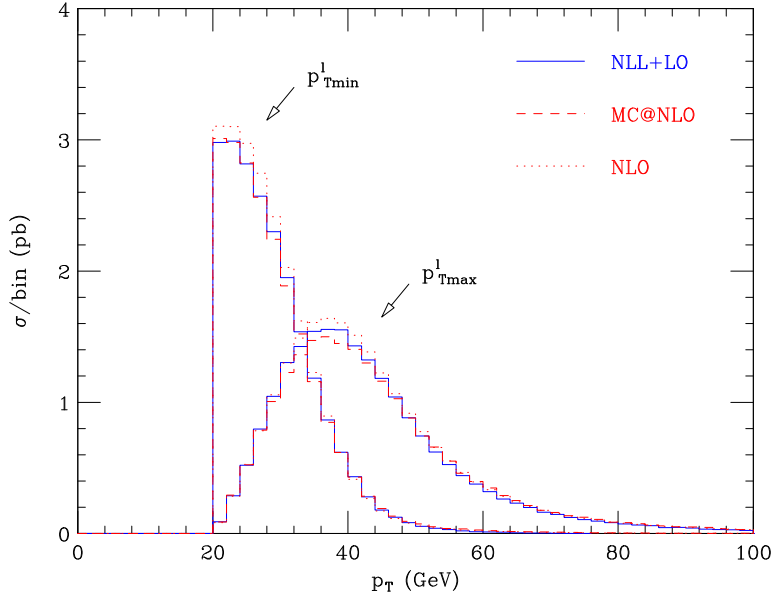}
\end{tabular}
\end{center}
\caption{\label{fig:ptleptnc}{\em Distributions in $\ptmin$ and $\ptmax$. Cuts A are applied.}}
\end{figure}

In Fig.~\ref{fig:ptleptnc} we show the distributions in $\ptmin$ and $\ptmax$:
as in Fig.~\ref{fig:ptlept}, no significant differences are found between the three histograms: these
distributions are still reliably predicted by the fixed-order NLO calculation.
This conclusion is confirmed by studying scale variations.
The effect of renormalization- and factorization-scale variations is still very small, and the impact of resummation-scale variations on the NLL+LO result
is again of about $\pm 2-3\%$ on average.

In the search for the Higgs boson in the $H\to WW\to l\nu l\nu$ channel
an important difference between the signal and the background is found in the $\Delta\phi$ distribution.
Since the Higgs is a scalar, the charged leptons tend to be produced quite close in angle.
As a consequence, the signal is expected to be peaked at small values of $\Delta\phi$,
whereas the $\Delta\phi$ distribution for the background is expected to be reasonably flat.
It is thus important to study the effect of resummation on this distribution,
which is also known to be particularly sensitive to spin correlations \cite{Gleisberg:2005qq}.

\begin{figure}[htb]
\begin{center}
\begin{tabular}{cc}
\epsfysize=8truecm
\epsffile{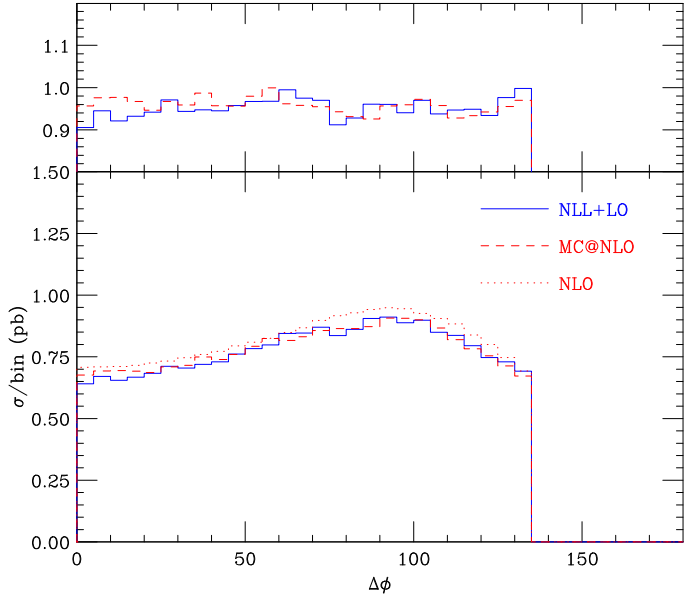}\\
\end{tabular}
\end{center}
\caption{\label{fig:dphinc}{\em The $\Delta\phi$ distribution when cuts A are applied. The upper part of the plot shows the NLL+LO and MC@NLO results normalized to the NLO one.}}
\end{figure}

In Fig.~\ref{fig:dphinc} the $\Delta\phi$ distribution at NLL+LO (solid) is compared
to the one at NLO (dotted) and from MC@NLO (dashes).
The upper panel shows NLL+LO and MC@NLO results normalized to the NLO prediction.
Note that both the NLO and the NLL+LO calculations fully include spin correlations,
whereas MC@NLO neglects spin correlations in the finite (non factorized)
part of the one-loop contribution.
We see that the agreement between the three results is excellent, showing that the approximate treatment of
spin correlations in MC@NLO is accurate enough.
Comparing NLL+LO and MC@NLO predictions with NLO ones we also see that the effects of resummation
appear negligible in this situation.
As in the case of Fig.~\ref{fig:ptleptnc}, this conclusion is confirmed by studying scale variations.
The effect of renormalization- and factorization-scale variations is still
of about $\pm 1\%$.
The effect of resummation-scale variations on the NLL+LO prediction,
not shown in the plot, is about $\pm 2\%$.

We now consider the application of stronger selection criteria \cite{Davatz:2004zg},
designed for the search of a Higgs boson with mass $M_H=165$ GeV. 

{\bf Cuts B}
\begin{itemize}
\item For each event, $\ptmin$ should be larger than $25$ GeV and $\ptmax$ should be between $35$ and $50$ GeV.
\item The invariant mass $m_{ll}$ of the charged leptons should be smaller than $35$ GeV.
\item The missing $p_T$ of the event should be larger than $20$ GeV.
\item The azimuthal separation $\Delta\phi$ of the charged leptons in the transverse plane should be smaller than $45^o$.
\item A jet veto is
mimicked
by imposing that the transverse momentum of the $\WW$ pair should be smaller than $30$ GeV.
This cut is perfectly legitimate in our resummed calculation and is exactly equivalent to a jet veto at NLO.
\end{itemize}
These cuts further select the small $\Delta\phi$ region. The jet veto is usually applied
in order to reduce
the $t{\bar t}$ contribution, which is expected to produce large-$p_T$ $b$-jets from the decay of the top quark.

The new cuts reduce the number of $\WW$ events by an additional factor of 35
with respect to cuts A.
The NLL+LO (MC@NLO) accepted cross section is 0.599 pb (0.570 pb) which should be contrasted with the NLO result,
which is 0.691 pb, about $20\%$ higher.
This relative large difference is due to the fact that the new cuts
enhance the relevance of the small-$\ptWW$ region,
where the NLO calculation is not any more reliable.

\begin{figure}[htb]
\begin{center}
\begin{tabular}{c}
\epsfysize=9truecm
\epsffile{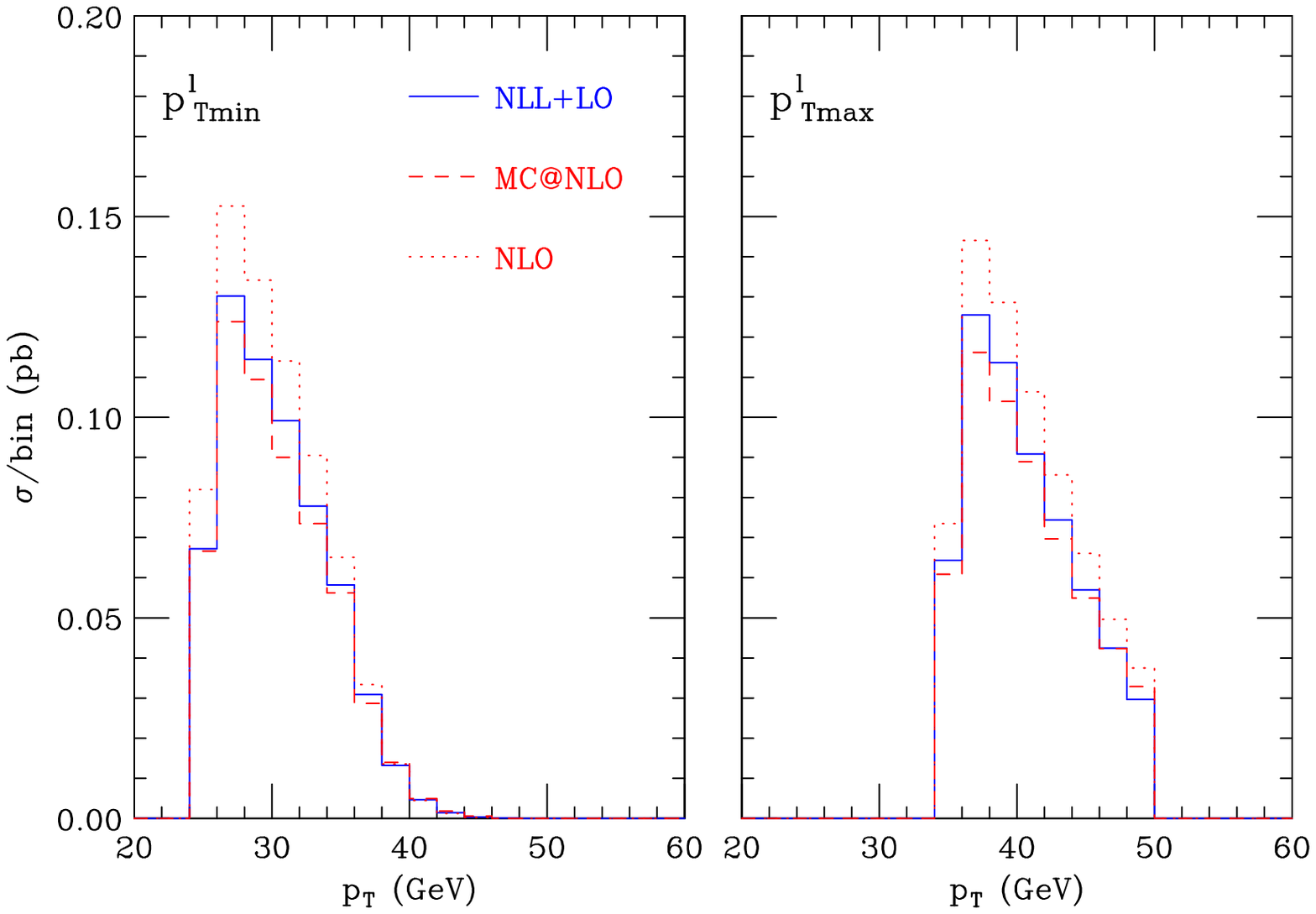}\\
\epsfysize=9truecm
\epsffile{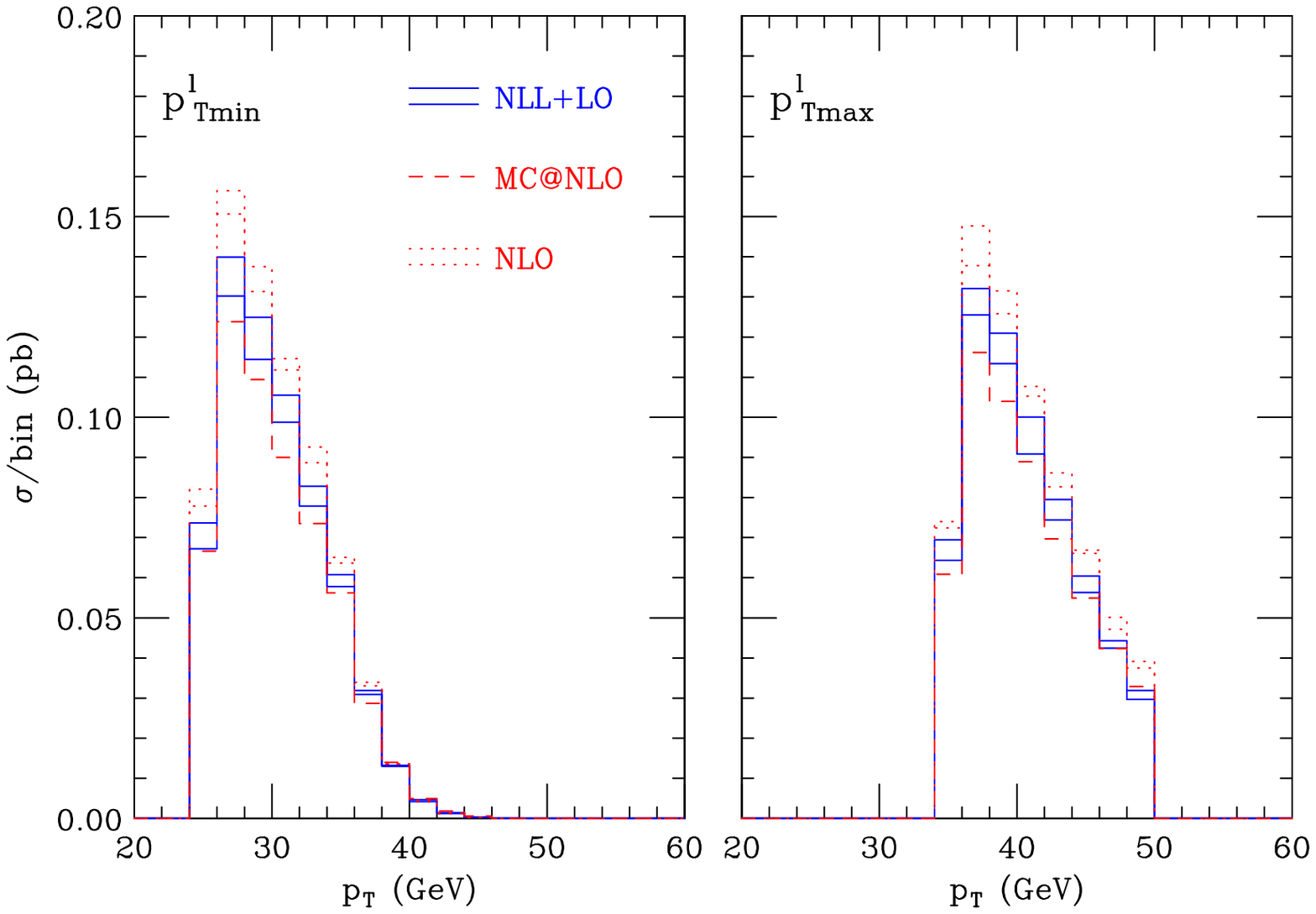}
\end{tabular}
\end{center}
\caption{\label{fig:ptlepthc}{\em Distributions in $\ptmin$ and $\ptmax$ when cuts B are applied.
In the lower plot the perturbative uncertainty bands at NLO and NLL+LO are shown.}}
\end{figure}

In Fig.~\ref{fig:ptlepthc} the $\ptmin$ and $\ptmax$ distributions are presented.
The upper plot shows the two spectra for central values of the scales. 
We see that although the three predictions are still
in reasonable agreement in shape,
differences are now evident.
In particular, the $\ptmin$ distribution at NLO is definitely steeper than the other two.
Comparing NLL+LO and MC@NLO spectra, we see that
the former are steeper than the latter:
with the application of stronger cuts
the differences between NLL+LO and MC@NLO predictions are clearly amplified.
The lower plot shows the uncertainty band
at NLO and NLL+LO. At NLO the band is obtained by varying $\mu_F=\mu_R$ between $M_W$ and $4M_W$:
the effect of scale variations is now more visible and is of the order of
a few percent.
The corresponding scale variations have marginal effect on the NLL+LO
and MC@NLO results:
this means that when the small $\ptWW$ region is selected, scale uncertainties increase at fixed order,
while remaining small when resummation is included.
At NLL+LO the band is obtained as before
by varying the resummation scale $Q$ between $M_{WW}$ and $4M_{WW}$, and the effect is of about $\pm 3-4\%$.

\begin{figure}[htb]
\begin{center}
\begin{tabular}{cc}
\epsfysize=6truecm
\epsffile{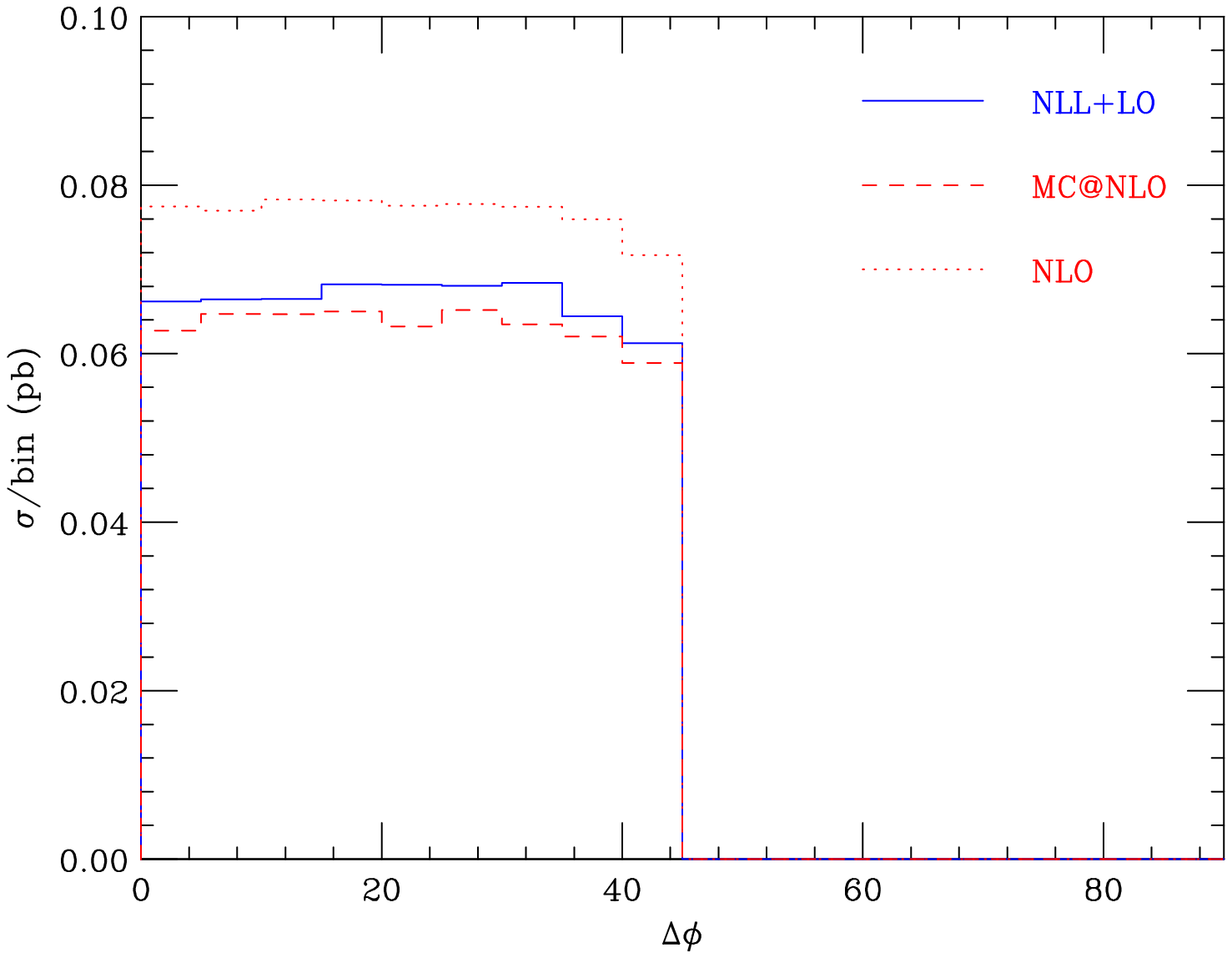} &\epsfysize=6truecm\epsffile{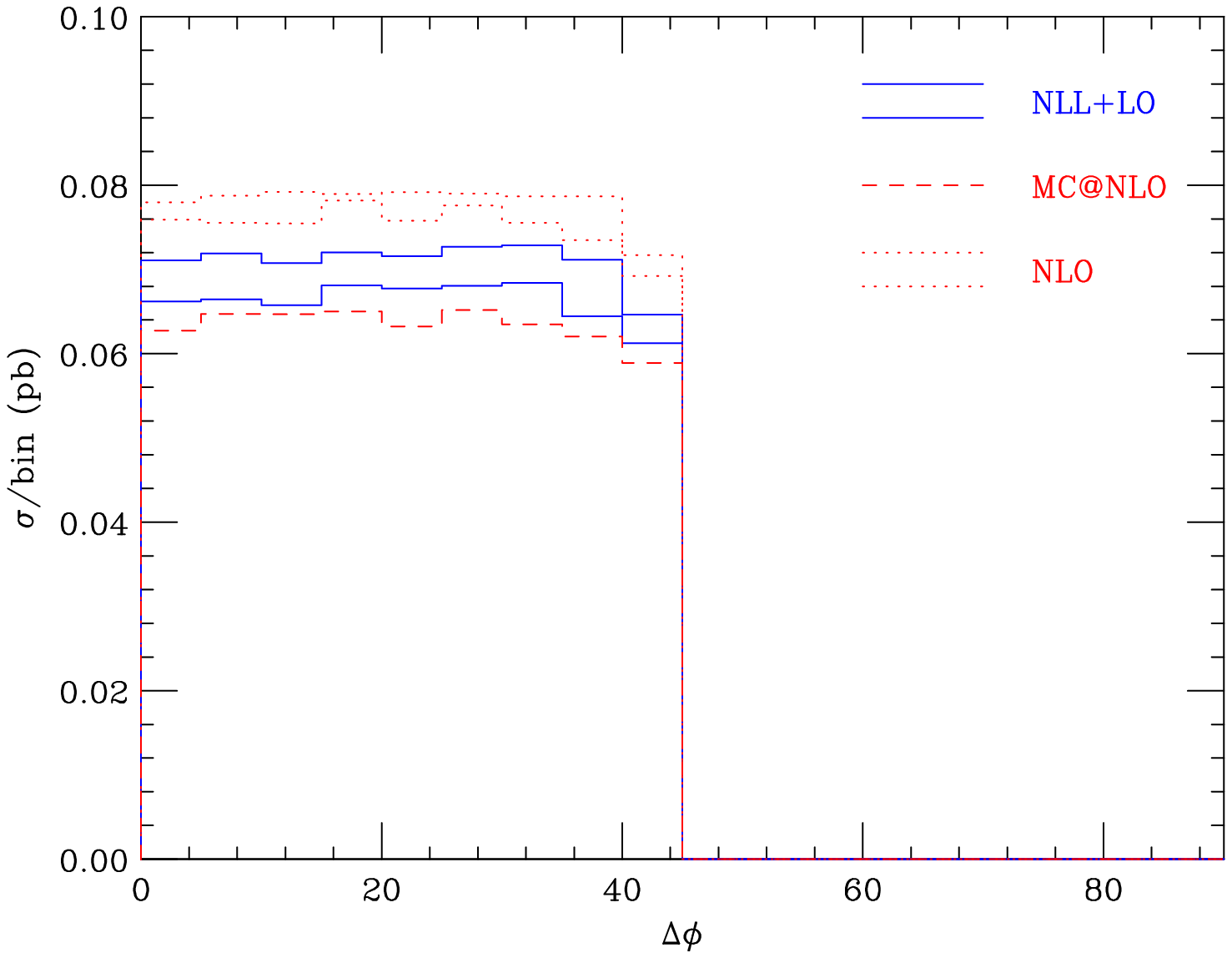}\\
\end{tabular}
\end{center}
\caption{\label{fig:dphihc}{\em Left: $\Delta\phi$ distribution when cuts B are applied.
Right: the corresponding NLL+LO and NLO uncertainty bands are shown. }}
\end{figure}
In Fig.~\ref{fig:dphihc} (left) the $\Delta\phi$ distribution is displayed.
The NLO and NLL+LO bands (right), computed as in Fig.~\ref{fig:ptlepthc}, are also shown.
We see that the shapes of the three results
are still in good agreement with each other, although
a slightly different slope of the
NLL+LO result with respect to MC@NLO and NLO ones starts to appear.

We finally consider the transverse-mass distribution of the $\WW$ system. We choose to define
the transverse mass according to Ref.~\cite{Rainwater:1999sd}.
We start from the transverse energy of the charged leptons and of the neutrinos, which are expressed as
\begin{equation}
E_{Tll}=\sqrt{p_{Tll}^2+m^2_{ll}}~~~~~~~~\slash E_{T}=\sqrt{\slash p_{T}^2+m^2_{\nu\nu}}\, ,
\end{equation}
where ${\bf p}_{Tll}$ and $\slash {\bf p}_T$ are the total transverse momentum of the charged leptons and the missing transverse momentum,
respectively.
If the Higgs boson mass is close to the $\WW$ threshold,
the invariant mass of the neutrinos in the expression of $\slash E_T$ can be approximated by $m_{ll}$ and
we can define
\begin{equation}
M_{TWW}=\sqrt{(\slash E_{T}+E_{Tll})^2-({\bf p}_{Tll}+\slash {\bf p}_T)^2}\, .
\end{equation}

\begin{figure}[htb]
\begin{center}
\begin{tabular}{cc}
\epsfysize=6truecm
\epsffile{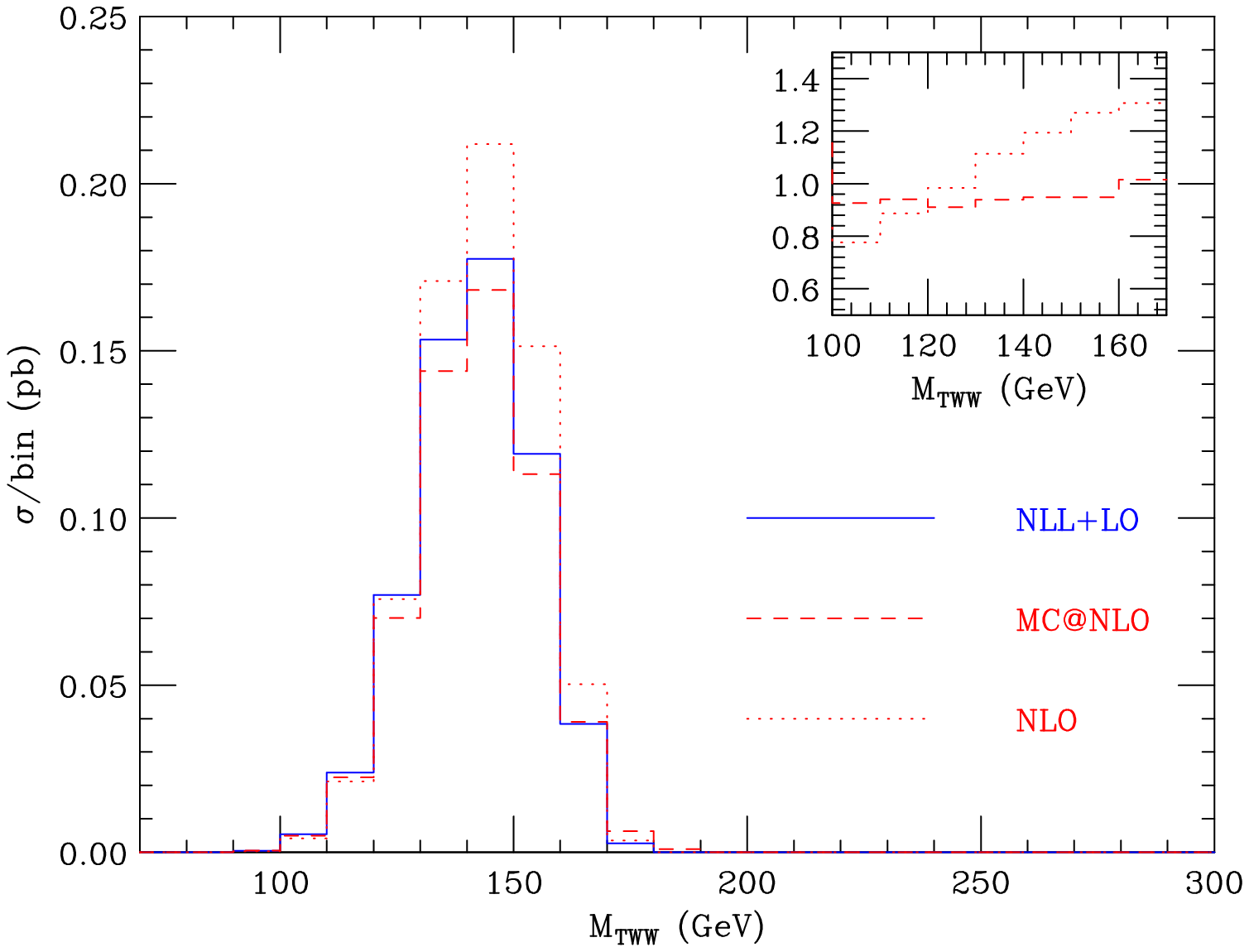} & \epsfysize=6truecm\epsffile{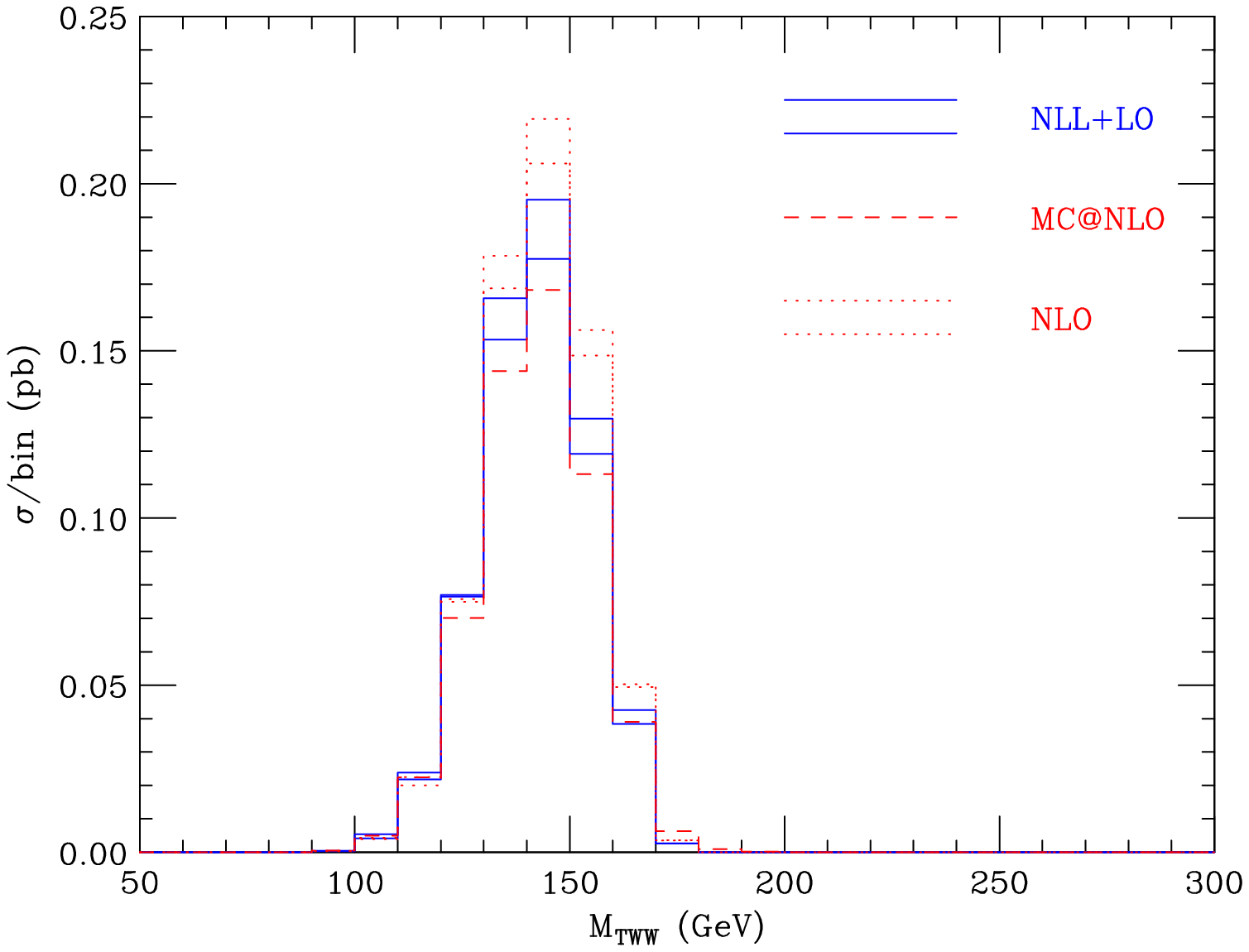}\\
\end{tabular}
\end{center}
\caption{\label{fig:mtww}{\em Left: $M_{TWW}$ distribution when cuts B are applied. The inset plot shows NLO and MC@NLO results normalized to NLL+LO. Right: NLL+LO and NLO uncertainty bands computed as in Fig.~\ref{fig:dphihc}.}}
\end{figure}
In Fig.~\ref{fig:mtww} (left)
we show the transverse-mass distribution defined above,
computed at NLO (dotted), NLL+LO (solid) and with MC@NLO (dashes). 
The inset plot shows the NLO and MC@NLO results normalized to NLL+LO.
The plot on the right shows the uncertainty bands computed
as in Figs.~\ref{fig:ptlepthc},\ref{fig:dphihc}.
All the three histograms are peaked at $M_{TWW}$ between 140 and 150 GeV.
Comparing the shapes of the histograms,
we see that at NLO the shape
is fairly different with respect to NLL+LO and MC@NLO.
The NLL+LO and MC@NLO distributions also show small differences:
the NLL+LO result is steeper and softer than the MC@NLO one.

\section{Summary}
\label{sec:concl}

In this paper we studied the effects of soft-gluon
resummation in $\WW$ pair production in hadron collisions.
We performed a calculation that, using the helicity amplitudes computed in Ref.~\cite{Dixon:1998py},
achieves uniform NLO accuracy over the whole phase space
but consistently includes the resummation of the large
logarithmic contributions at $\ptWW\ll M_{WW}$. 
We presented predictions for the $\ptWW$ spectrum
at NLL+LO and (almost) NNLL+LO accuracy, and compared our results with those
obtained with the MC@NLO event generator, finding good agreement.

We then examined a few charged-lepton distributions:
the $p_T$ spectrum of the lepton with minimum and maximum $p_T$,
the azimuthal separation of the charged leptons in the transverse plane and the
transverse-mass distribution, when different sets of cuts are applied.
The $p_T$ spectra of the charged leptons are generally well described by the NLO calculation
but the effect of resummation becomes visible when hard cuts are applied.
The $\Delta\phi$ distribution is
important to discriminate a Higgs-boson signal over the background
in the $H\to WW\to l\nu l\nu$ channel,
and is also known to be particularly sensitive to the effect of spin correlations
in the $W$'s decay. The latter are fully taken into account at NLO and in our NLL+LO calculation,
whereas MC@NLO neglects spin correlations in the finite (non-factorized) part of the one-loop contribution.
All the predictions for the $\Delta\phi$ distribution
agree well and resummation effects appear small even when hard cuts are applied.
We also examined
the transverse mass distribution of the WW pair.
The shape of the NLO distribution differs substantially from NLL+LO and MC@NLO.
On the other hand, the NLL+LO and MC@NLO results are still
in reasonable agreement.

We finally note that the resummed predictions presented
in this paper were obtained in a purely
perturbative framework.
Two kinds of non-perturbative effects may be considered: those due to hadronization, and those
due to an intrinsic transverse momentum (intrinsic $p_T$) of the partons in the colliding hadrons.

In the present paper we were mainly concerned with leptonic observables.
As a consequence, we do not expect hadronization effects to be particularly relevant.
This expectation is supported by
the good agreement we find between resummed results
and MC@NLO predictions, in which hadronization effects
{\em are} taken into account.

As far as intrinsic-$p_T$ effects are concerned,
it is known (see e.g. Ref.~\cite{Collins:va} 
and references therein)
that transverse-momentum distributions
are affected by such kind of non-perturbative effects,
particularly
at small transverse momenta.
These effects,
associated to the large-$b$ region in impact parameter,
are not taken into account in our calculation.
As noted in Ref.~\cite{Bozzi:2005wk},
these non-perturbative effects
have the same qualitative impact
as the inclusion of higher-order logarithmic contributions, i.e., they tend
to make the $p_T$ distribution harder.
The precise quantification of non-perturbative effects in $WW$ production is left for future investigations.

\noindent {\bf Acknowledgements}

\noindent I wish to thank Stefano Catani for valuable discussions and comments.
I also thank Daniel de Florian, Gunther Dissertori,
Michael Dittmar, Volker Drollinger, Stefano Frixione, and Zoltan Kunszt
for helpful discussions
and suggestions, and CERN Theory Unit for the
hospitality and financial support at various
stages of this work.

\end{document}